# Impact of Zn substitution on phase formation and superconductivity of $Bi_{1.6}Pb_{0.4}Sr_2Ca_2Cu_{3-x}Zn_xO_{10-\delta}$ with x = 0.0, 0.015, 0.03, 0.06, 0.09 and 0.12.

R.B. Saxena[1], Rajiv Giri[2], V.P.S. Awana[1,*], H.K. Singh[1], M.A. Ansari[1], B.V. Kumaraswamy[1], Anurag Gupta[1], Rashmi Nigam[1], K.P. Singh[1], H. Kishan[1] and O.N. Srivastava[2]

[1]National Physical Laboratory, K.S. Krishnan Marg, New Delhi 110012, India.

[2] Department of Physics, Banaras Hindu University, Varanasi 221005, India.

Samples of series $Bi_{1.6}Pb_{0.4}Sr_2Ca_2Cu_{3-x}Zn_xO_{10-\delta}$ with x = 0.0, 0.015, 0.03, 0.06, 0.09 and 0.12 are synthesized by solid-state reaction route. All the samples crystallize in tetragonal structure with majority (> 90%) of Bi-2223 ($Bi_2Sr_2Ca_2Cu_3O_{10}$) phase ($c$-lattice parameter ~ 36 $A^0$). The proportion of Bi-2223 phase decreases slightly with an increase in x. The lattice parameters $a$ and $c$ of main phase (Bi-2223) do not change significantly with increasing x. Superconducting critical transition temperature ($T_c$) decreases with x as evidenced by both resistivity [$\rho(T)$] and AC magnetic susceptibility [$\chi(T)$] measurements. Interestingly the decrement of $T_c$ is not monotonic and the same saturates at around 96 K for x > 0.06. In fact $T_c$ decreases fast (~10K/at%) for x = 0.015 and 0.03 samples and later nearly saturates for higher x values. Present results of Zn doping in Bi-2223 system are compared with Zn doped other HTSC (High temperature superconducting) systems, namely the RE-123 ($REBa_2Cu_3O_7$) and La-214 ($(La,Sr)_2CuO_4$).

Key Words: Bi-2223, Zn doping effects, AC susceptibility, Resistivity, XRD

*: Corresponding Author: E-mail; awana@mail.nplindia.ernet.in

**INTRODUCTION**

In highly an-isotropic layered HTSC (High temperature superconducting) compounds, the superconductivity is supposed to reside in $Cu-O_2$ planes, which are integral part of all these systems [1,2]. Except one HTSc family i.e. $REBa_2Cu_3O_7$ (RE = rare earth), which has two Cu-sites in its structure namely; (a) Cu(I) in $Cu-O_{1-\delta}$ chains and (b) Cu(II) in $Cu-O_2$ planes, all others have only one site i.e. (b). Any alteration in the $Cu-O_2$ planes in terms of disorder etc., results in drastic reduction of superconductivity [1-3]. In this regards, Cu-site partial substitution of $Zn^{2+}$ has attracted a lot of attention since from the invention of HTSC [3-6]. The size of square planer Cu in $Cu-O_2$ planes marches well with that of $Zn^{2+}$ hence the partial substitution can take place without much changing the lattice parameters of the doped system [3-7].

Cu in $Cu-O_2$ planes can be partially substituted easily by Ni and Zn due to their closer ionic size proximity. Zn substitution in HTSC was considered to be anomalous and most interesting in the beginning of HTSC invention itself, in particular due to the reason that the substitution of magnetic divalent ion Ni at Cu-site showed five times smaller $T_c$ depression rate than that of non-magnetic ion $Zn^{2+}$ at the same site [4-8]. It was noted that non magnetic ion $Zn^{2+}$ introduces a local magnetic moment of nearly $0.70\mu_B$ supposedly on Cu, which breaks the superconducting pairs much stronger than a widely distributed $3.2\mu_B$ moment of $Ni^{2+}$ ions in $Cu-O_2$ planes [9-11]. Understanding of Ni and Zn metals substitutions at Cu-site in HTSC compounds was further complicated by the measurements on doped Nd-214 and RE-124 systems [12-15]. In these compounds, dopings with Ni and Zn have shown similar $T_c$ depression rates with an increase in their doping concentration in the pristine system. The results of Nd-214 were interpreted on the basis of larger coherence length of this compound in ab-plane (90 Å), than for RE-123 and La-214 compounds. The mystery of Ni and Zn substitution at Cu-site in RE-124 system is still open, despite some trials [15]. In another analysis den Hertog and Das have shown that Zn doping in cuprates induces a magnetic moment near the impurity site. They argue that Zn does not necessarily have a valency of 2+. Its valency is determined by surrounding correlation medium. Therefore it is not correct to assign a valency to Zn or Ni before their substitution for Cu in the cuprate system [16].

Interestingly enough the substitution of Zn at Cu-site in Bi, Tl or Hg based cuprates is not explored, only few scant reports lie in the literature [17-20]. This is important with the fact that many of these systems do have more than two Cu-O$_2$ planes in their structure, unlike widely studied Zn substituted RE-123 system. It is also yet debatable that whether superconductivity resides in all Cu-O$_2$ planes or only in selected ones. Few reports yet available on Zn doping in Bi-2223 have resulted in stabilization of low T$_c$ (85 K) Bi-2212 (Bi$_2$Sr$_2$CaCu$_2$O$_8$) phase and saturation of T$_c$ at around 90 K for Cu/Zn substitution of around more than 1at% [19,20].

In the present letter in order to revisit the Zn substitution in Bi-2223 system, we carefully synthesized these compounds and characterized them for phase formation, resistivity and susceptibility measurements. The T$_c$ (107 K) of pristine Bi-2223 sample decreased to 98 K for 1at% Zn/Cu substitution. Later this value is saturated and T$_c$ does not decrease further with substitution. Surprisingly in our samples the Bi-2212 phase formation is not as much as reported earlier. We speculate that after some level of doping Zn does not substitute at Cu in Cu-O$_2$ planes and hence not much deleterious effect on T$_c$.

## EXPERIMENTAL DETAILS

Samples of series Bi$_{1.6}$Pb$_{0.4}$Sr$_2$Ca$_2$Cu$_{3-x}$Zn$_x$O$_{10-\delta}$ (x = 0.0, 0.03, 0.06, 0.09 and 0.12) are synthesized by solid-state reaction route with ingredients of Bi$_2$O$_3$, Pb$_2$O$_3$, CaCO$_3$, SrCO$_3$, CuO and Zn O of having more than 4N (99.99) purity. Calcinations were carried out on mixed powders at 800°C, 810°C, and 920°C each for 24 hours with intermediate grindings. Pressed pellets were annealed in a air at 840°C for 300 hours and subsequently cooled slowly to room temperature X-ray diffraction (XRD) patterns were obtained at room temperature with Cu$K_\alpha$ radiation. Four probe resistance measurements were carried out on a close cycle refrigerator from 300 K down to 12 K. The system is hooked to a computer and all the data acquisition is automatic. AC magnetic susceptibility measurements are carried out on a Lakeshore AC susceptometer (model AC7000) down to 77 K. The frequency and amplitude used for AC magnetic field are 111.1 Hz and 200 Amp/m for all the samples.

## RESULTS AND DISCUSSION

The Structural phase characterization was carried out by powder X-ray diffraction employing Cu K$_\alpha$ radiation and the XRD data corresponding to all the samples are plotted in figure 1. It is seen in the diffraction pattern that Zn doping only has little effect on reflection-intensity profile as a function of the angle of diffraction. The XRD data has been indexed using an orthorhombic unit cell and the most intense reflections appearing in the XRD spectra of each sample are listed in table 1.

**Table 1.**

| Reflection # | x =0.0 | x=0.03 | x=0.06 | x=0.09 | x=0.12 |
|:---:|:---:|:---:|:---:|:---:|:---:|
| 1 | (200) | (00$\underline{12}$) | (00$\underline{12}$) | (00$\underline{12}$) | (200) |
| 2 | (00$\underline{12}$) | (200) | (00$\underline{14}$) | (200) | (00$\underline{12}$) |
| 3 | (220) | (00$\underline{10}$) | (00$\underline{10}$) | (00$\underline{14}$) | (00$\underline{10}$) |
| 4 | (11$\underline{11}$) | (119) | (200) | (00$\underline{10}$) | (00$\underline{14}$) |
| 5 | (02$\underline{12}$) | (00$\underline{14}$) | (11$\underline{11}$) | (119) | (11$\underline{11}$) |

The appearance of strong reflections corresponding to the (002), (00$\underline{12}$) and (00$\underline{14}$) planes of Bi-2223 phase having respective d-values as $d_{002} = 18.567$ Å, $d_{00\underline{12}} = 3.095$ Å and $d_{00\underline{14}} = 2.652$ Å and simultaneous absence of any such reflection corresponding to the lower Bi-2212 phase suggests that our samples are dominantly 2223. It must be mentioned that the lower phase 2212 does not have the any reflection with above mentioned d-values. Further, some more strong and unambiguous reflections corresponding to the 2223 phase such as (11$\underline{11}$) and (11$\underline{19}$) having d-values $d_{11\underline{11}} = 2.535$ Å and $d_{11\underline{19}} = 1.741$ Å are also observed in the XRD data. No significant change in the XRD spectra and hence the lattice parameters have been observed as a function of the Zn concentration. The lattice parameters of the orthorhombic unit cell calculated using the XRD data have been found to be a = 5.407 ± 0.005 Å, b = 5.415 ± 0.007 Å and 37.135 ± 0.008 Å. The limits specified above are valid for all the Zn concentrations and this shows that the lattice parameters are nearly independent of Zn concentration. No trace of any independent presence of Zn or its oxide has been found in the XRD data. Thus on the basis of our XRD data it can be concluded that all the samples are indeed nearly single phase 2223.

The superconducting transition was determined by both, ac susceptibility as well as the four probe dc resistance measurements. The temperature dependence of ac susceptibility and the resistance are plotted in figure 2 and figure 3 respectively. The $T_c$ values as measured by ac susceptibility are found to be 108 K, 102 K, 99 K, 98 K, 98 K and 98 K respectively for Zn concentrations x = 0.0, 0.015, 0.03, 0.06, 0.09 and 0.12. The dc transport measurements reveal the onset of transition at 110 K, 107 K, 105 K. 104 K, 104 K and 104 K for for Zn concentrations x = 0.0, 0.015, 0.03, 0.06, 0.09 and 0.12 and the corresponding $T_c$ (R=0) occurs at 107 K, 102 K, 98 K, 95 K, 95 K and 95 K. Thus the transition width, $\Delta T$, corresponding to Zn concentrations x = 0.0, 0.015, 0.03, 0.06, 0.09 and 0.12 are 3 K, 7 K, 9 K, 9 K and 9 K respectively. The magnified view of the R-T data around transition is plotted in figure 4 while the dependence of $T_c$ (onset), $T_c$ (R=0) and $\Delta T$ are plotted in figure 5. Since there is little change in the structural phase profile as a function of Zn concentration, the observed trend in the transport properties cannot be attributed to any additional phase growth due to Zn doping This observed $T_c$ variation where there is a sudden strong decrement in $T_c$ with Zn doping x ≤ 0.06 and beyond this (x > 0.06) there is saturation in all the parameters, viz., $T_c$(onset), $T_c$(R=0) and the transition width $\Delta T$ it as a function of Zn concentration is in sharp contrast to the previously observed results where in most of the cases a linear depression in $T_c$ with Zn concentration is observed. Thus in Bi-bearing 2212 HTSCs $dT_c/dx$ in the range -500 K to -1800 K have been reported [1-5]. The Bi-2223 system is bit more complex than the 2212 (where the two Cu-O sheets are equivalent for a dopant such as Zn, Ni, Co etc.) in the sense that there are two slightly different types of Cu sites, one on the central of the three closely spaced Cu-O sheets and the other on any one of the two outer Cu-O sheets, where an impurity such as Zn can get substituted. It has been shown that superconductivity lies on the central Cu-O plane and any Zn substitution on this plane only will affect the $T_c$ of the system. Thus one of the reasons for the saturation of the $T_c$ seems to be the limited solubility of Zn in the central Cu-O planes that are critical to the occurrence of superconductivity. On the basis of our magneto-transport measurement the observed trends in $T_c$ can be accounted for by assuming that the solubility of dopant Zn in the central Cu-O plane may be limited and that beyond a critical fraction the remainder Zn atoms are substituted in the two outer Cu-O planes. Thus initially (x ≤ 0.06) the Zn cations are substituted at the Cu sites on the central Cu-O planes and consequently induce a localized magnetic moment on Cu resulting in the destruction of the superconducting pairs. Further, since $Zn^{2+}$ is larger than the $Cu^{2+}$ cation the disorder due the size mismatch may also have some bearing on the $T_c$ and contribute to observed $T_c$ decrement. As the Zn concentration is increased the additional Zn atoms are substituted at the Cu sites in the two outer Cu-O planes in order to minimize the disorder in the central Cu-O plane. This constraint of limited doping in the central Cu-O plane where the superconductivity lies in tri Cu-O materials such as Bi-2223, Tl-2223, Tl-1223 and Hg-1223 may limit the decrements in the $T_c$.

**CONCLUSION**

In conclusion, we had synthesized nearly phase pure samples of series $Bi_{1.6}Pb_{0.4}Sr_2Ca_2Cu_{3-x}Zn_xO_{10-\delta}$ with x = 0.0, 0.015, 0.03, 0.06, 0.09 and 0.12 having more than 90% Bi-2223 phase. The lattice parameters *a* and *c* of main phase (Bi-2223) are nearly unchanged with x. $T_c$ decreases with x as evidenced by both resistivity [$\rho(T)$] and AC magnetic susceptibility [$\chi(T)$] measurements. The decrement of $T_c$ is not monotonic and the same saturates at around 96 K for x > 0.06. The results are compared with Zn doped other HTSC systems.

**ACKNOWLEDGEMENT**

Authors from NPL, thank Prof. Vikram kumar Director for his keen interest in the present work.

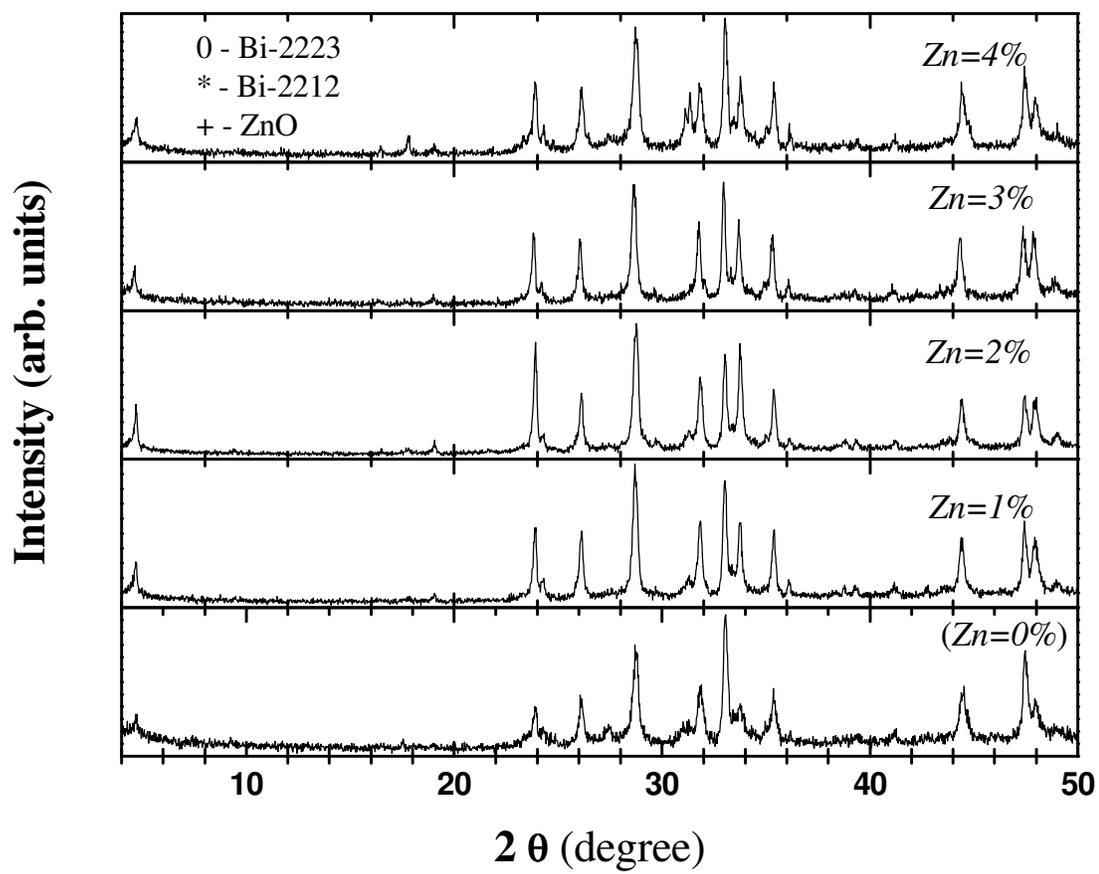

Figure 1, Saxena et al.

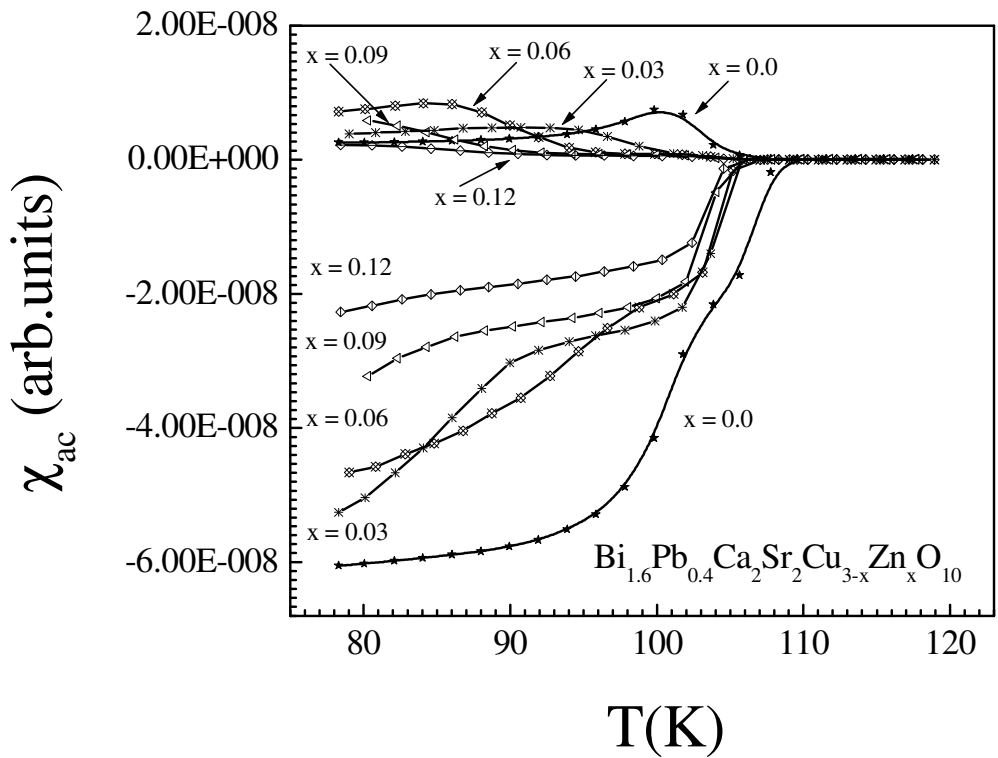

Figure 2 Saxena et al.

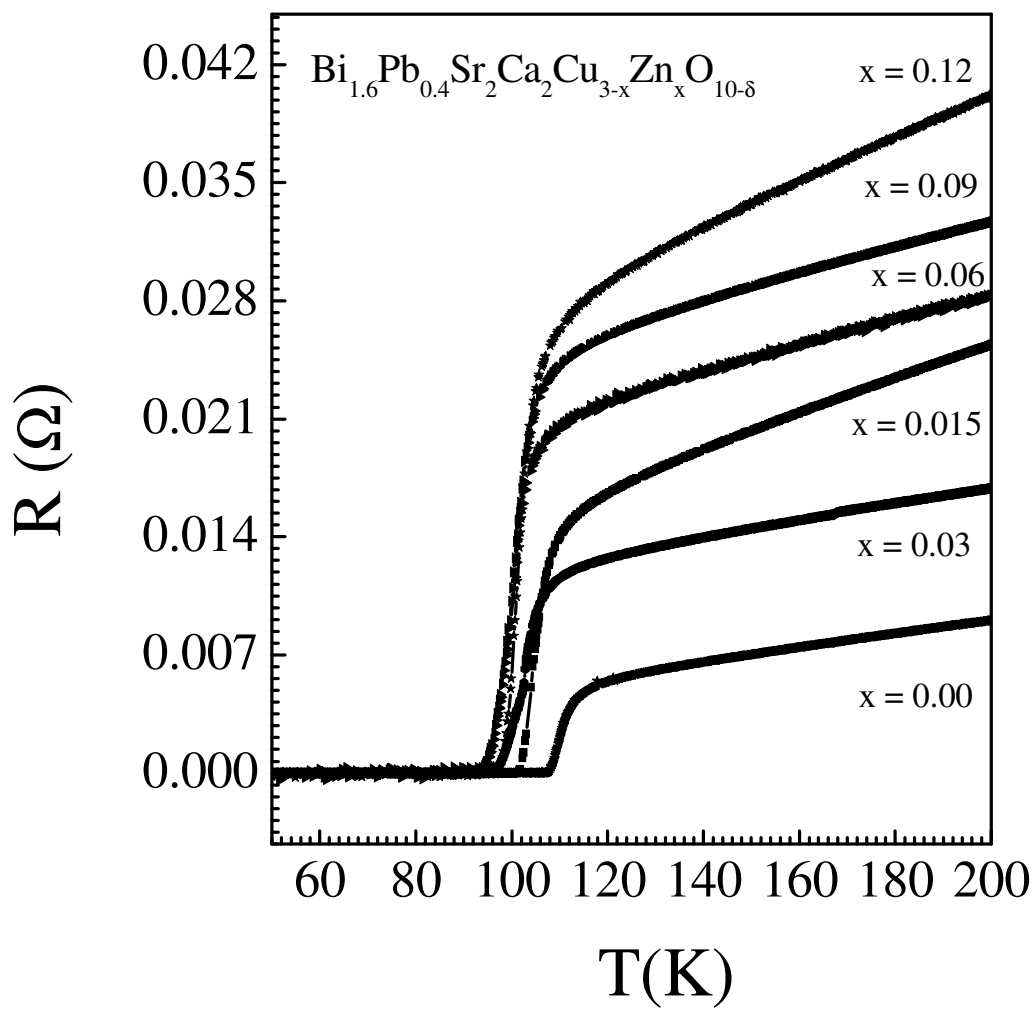

$Bi_{1.6}Pb_{0.4}Sr_2Ca_2Cu_{3-x}Zn_xO_{10-\delta}$

x = 0.12

x = 0.09

x = 0.06

x = 0.015

x = 0.03

x = 0.00

R ($\Omega$)

T(K)

Figure 3 Saxena et al.

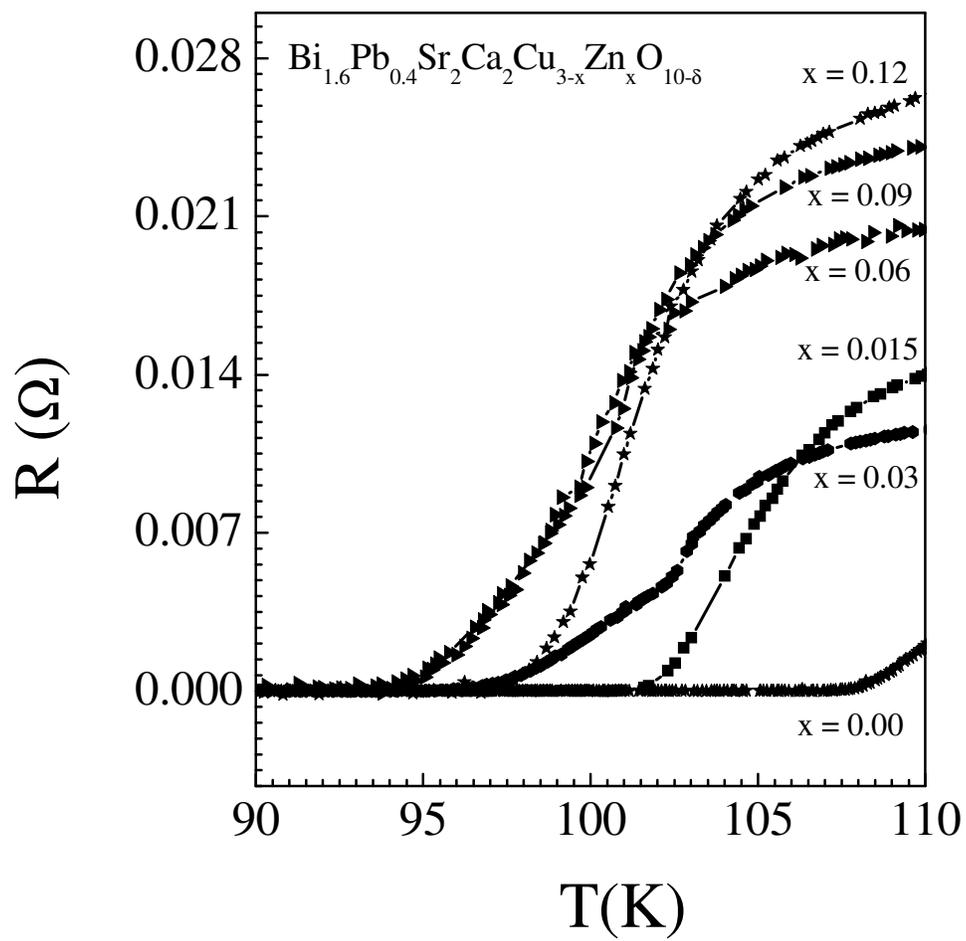

Figure 4 Saxena et al.

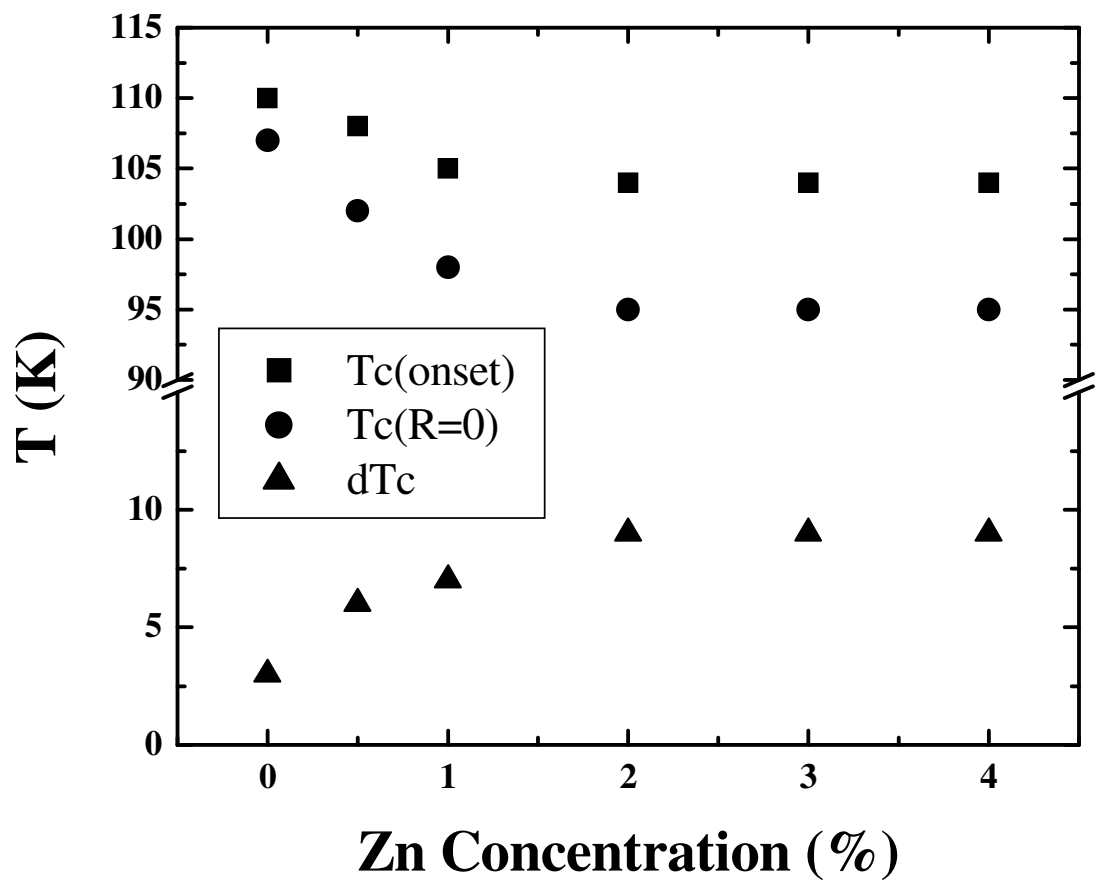

Figure 5 Saxena et al.